%% file: main.tex
\documentclass[11pt]{article}

\usepackage[margin=1in]{geometry}
\usepackage{amsmath,amssymb,amsthm,mathtools}
\usepackage{graphicx,tikz}
\usepackage{enumitem}
\usepackage{booktabs}
\usepackage{float}
\usepackage{microtype}
\usepackage{xcolor, xspace}
\usepackage{thmtools}
\usepackage{thm-restate}
\usepackage{algorithm}
\usepackage{algpseudocode}
\usepackage{hyperref}
\usepackage[nameinlink,capitalize,noabbrev]{cleveref}

\newtheorem{theorem}{Theorem}[section]
\newtheorem{lemma}[theorem]{Lemma}
\newtheorem{proposition}[theorem]{Proposition}

\newtheorem{claim}[theorem]{Claim}
\theoremstyle{definition}
\newtheorem{definition}[theorem]{Definition}
\newtheorem{remark}[theorem]{Remark}

\newcommand{\pr}{\text{Pr}}
\newcommand{\R}{\mathbb{R}}
\newcommand{\Z}{\mathbb{Z}}
\newcommand{\calF}{\mathcal{F}}
\newcommand{\calA}{\mathcal{A}}
\newcommand{\calB}{\mathcal{B}}
\newcommand{\calC}{\mathcal{C}}

\newcommand{\conv}{\text{conv}}
\newcommand{\proj}{\text{Proj}}

\newcommand{\pEC}{\textsc{$p$-Edge-Cover}\xspace}
\newcommand{\EC}{\textsc{Edge-Cover}\xspace}
\newcommand{\FEC}{\textsc{Flex-Edge-Cover}\xspace}
\newcommand{\pqFGC}{$(p,q)$-FGC\xspace}

\newcommand{\FGC}{\textsc{Flex-Graph-Connectivity}\xspace}
\newcommand{\DEC}{\textsc{Demand-Edge-Cover}\xspace}
\newcommand{\VC}{\textsc{Vertex-Cover}\xspace}
\newcommand{\MCUW}{\textsc{Disjunctive-Knapsack-Cover}\xspace}

\def\final{0}  
\def\iflong{\iffalse}
\ifnum\final=0  
\newcommand{\knote}[1]{{\color{blue}[{Karthik: \bf #1}]\marginpar{\color{blue}*}}}
\newcommand{\todo}[1]{{\color{red}[{TODO: \bf #1}]\marginpar{\color{red}*}}}
\newcommand{\krnote}[1]{{\color{purple}[{Krishna: \bf #1}]\marginpar{\color{purple}*}}}
\else 
\newcommand{\knote}[1]{}
\newcommand{\todo}[1]{}
\newcommand{\krnote}[1]{}
\fi  

\title{
Flexible Edge Cover: A Disjunctive Programming driven Approximation\thanks{Grainger College of Engineering, University of Illinois, Urbana-Champaign, Email: {\tt\{karthe, kk17\}@illinois.edu}. Supported in part by NSF grant CCF-2402667. }}
\author{Karthekeyan Chandrasekaran \and Krishna Kalathur}
\date{}

\begin{document}
\maketitle
\input{abstract}
\input{intro-2}
\input{Hardness}
\input{Algorithms-2}

\input{FGC_Disjunctive-2}

\bibliographystyle{abbrv}
\bibliography{references}

\input{appendix}

\end{document}

%% file: abstract.tex
\begin{abstract}
    We introduce the flexible edge cover problem, denoted \FEC: the input here is an undirected graph with non-negative edge costs along with a partition of the edges into safe and unsafe edges, non-negative integers $p$ and $q$, and the goal is to find a minimum cost subset of edges such that each vertex has either at least $p$ safe edges incident to it or at least $p+q$ total edges incident to it. We show that \FEC is APX-hard in simple bipartite graphs. A greedy approach leads to a $2$-approximation in all graphs. We show that the $2$-factor can be improved in simple graphs via a novel disjunctive programming-based LP formulation. AI-tools were used in the analysis of our rounding algorithm. 
    
    Our motivation for introducing and investigating \FEC is its close connection to the Flexible Graph Connectivity problem (denoted \pqFGC) which has garnered much recent attention. For \pqFGC, we develop a tighter LP-formulation via the disjunctive programming approach. We hope that our disjunctive-programming based LP and rounding algorithms for \FEC will serve as a starting point towards a constant factor approximation for \pqFGC.
\end{abstract}

%% file: intro-2.tex
\section{Introduction}
\EC is a classic problem in combinatorial optimization that aims to find a minimum cost network so that no node is isolated. It is a fundamental model for local coverage in network design. 
Formally, given a graph $G=(V, E)$ with non-negative costs $c: E\rightarrow \R_{\ge 0}$, the goal is to find a minimum cost subset $F\subseteq E$ of edges such that every vertex is incident to at least one edge of $F$. Its natural generalization, denoted \pEC, requires a minimum cost subset $F$ of edges such that every vertex is incident to at least $p$ edges in $F$. Both problems are solvable in polynomial time via reductions to maximum weight matching and maximum weight $b$-matching, respectively \cite{schrijver-book}. 

Motivated by the varying reliability of network components, a substantial body of work on network design has developed and studied non-uniform failure models, including flexible network design, bulk-robust network design, and relative survivable network design \cite{ASZ15, CJ2025, DKK22}. The same considerations arise for local coverage. For example, protected wired links may remain operational while wireless links could be disrupted by radio interference. Preserving the required number of operational links at every site creates a tradeoff between investing in protected connections and providing additional redundancy for vulnerable ones. 

We introduce \FEC to capture this tradeoff. The input is an undirected graph $G=(V, E=S\uplus U)$, whose edges are partitioned into safe edges $S$ and unsafe edges $U$, non-negative edge costs $c: E\rightarrow \Z_{\ge 0}$, and non-negative integers $p, q\ge 0$. For a vertex $v\in V$ and a set $F$ of edges, we write $\delta_F(v)$ to denote the edges of $F$ incident to vertex $v\in V$ and 
say that $v$ is \emph{safe-covered} by $F$ if $|\delta_{F\cap S}(v)|\ge p$, $v$ is \emph{cumulative-covered} by $F$ if $|\delta_F(v)|\ge p+q$, and $v$ is \emph{$(p,q)$-Flex-Covered} by $F$ if $v$ is either safe-covered or cumulative-covered by $F$. 
The goal is to find a minimum cost subset $F\subseteq E$ such that every $v$ is $(p, q)$-Flex-Covered, i.e., $F$ satisfies 
\begin{align}
|\delta_{F\cap S}(v)|\ge p & \quad \quad \quad \text{ or }\quad \quad \quad  |\delta_F(v)|\ge p+q\quad \quad \quad \quad \text{ for every }v\in V. \label{alternatives}
\end{align}
Equivalently, we are seeking a minimum-cost subset $F$ of edges such that $F$ remains a $p$-edge-cover after the failure of any set of at most $q$ unsafe edges. We allow parallel edges unless otherwise stated. 

Our motivation for \FEC also comes from the flexible edge connectivity problem, denoted \pqFGC, which requires a minimum cost subgraph that remains $p$-edge-connected after the failure of any set of at most $q$ unsafe edges. Introduced for $(p,q)=(1,1)$ by Adjiashvili, Hommelsheim, and M\"uhlenthaler \cite{AHM2022} and generalized by Boyd, Cheriyan, Haddadan, and Ibrahimpur \cite{BCHI24}, this problem equivalently imposes the same alternative as \eqref{alternatives} on every non-empty proper cut: at least $p$ safe edges or at least $p+q$ total edges should cross every non-empty proper cut. Thus, \FEC is a relaxation of \pqFGC since it imposes requirements only across singleton cuts. 
Several recent works including that of Chekuri-Jain \cite{CJ2025} and Ibrahimpur-V\'{e}gh \cite{IV25} developed LP-formulations for \pqFGC via combinatorial arguments. 
Despite substantial progress on this problem \cite{BCGI23,Ban25,BCKS25,BCHI24,CJ2025}, including an $O(\log{|V|})$-approximation \cite{IV25} and a very recent $(p+q)^{O(pq)}$-approximation \cite{CJK26}, a universal constant factor approximation independent of $p$ and $q$ remains open. Our interests in \FEC originated from here. \FEC localizes the disjunctive requirements to individual vertices, providing a setting in which to develop relaxations and rounding methods that might generalize to the more general \pqFGC.

From an integer programming perspective, \FEC is a covering problem with an explicit disjunction at each vertex. Optimizing at each vertex in isolation is easy: choose the cheaper of its $p$ cheapest safe incident edges and its $p+q$ cheapest incident edges. Taking the union of these local choices gives a $2$-approximation. Improving on this bound requires exploiting the fact that an edge serves both end-vertices. A natural LP with a fractional variable selecting each vertex's alternative has unbounded integrality gap---see Section \ref{sec:lp1-integ-gap}. 
We observe that the disjunctive constraint structure at each vertex lends itself to the use of disjunctive programming principles developed by Balas. 
We use Balas' ideas to formulate a stronger LP by convexifying each local feasible set exactly and coupling these convex hulls through shared edge variables. 
We design a rounding algorithm relative to this stronger LP to improve on the $2$-approximation for simple graphs. We complement our approximation results by showing that the problem is APX-hard on simple bipartite graphs. 

Our disjunctive programming approach also extends to \pqFGC leading to a new LP-relaxation. We show that this relaxation is at least as strong as those of Chekuri-Jain \cite{CJ2025} and Ibrahimpur-V\'{e}gh \cite{IV25}. We observe that this relaxation admits an efficient separation oracle and is hence, polynomial-time solvable via the Ellipsoid method. We conjecture that its integrality gap is bounded by a universal constant independent of $p$ and $q$. 
More broadly, our work advances the use of disjunctive programming in approximation algorithms by translating disjunctive constraints into improved LP-relative guarantees, providing a stepping stone toward a constant-factor LP-rounding algorithm for \pqFGC. 

\subsection{Results}
If $q=0$, then \FEC is reduces to the polynomial-time solvable \pEC. However, for general $q$, the problem is APX-hard (even on simple bipartite graphs for $p=1$ and $q=2$). 


\begin{theorem}\label{thm:hardness}
    For every $\epsilon > 0$, \FEC is NP-hard to approximate within a factor of \item $\frac{496}{495}-\epsilon$ on simple bipartite graphs even for $p=1$ and $q=2$.
\end{theorem}

The local $2$-approximation described earlier has a tight example---see Section \ref{sec:greedy-tight-example}. Our main algorithmic result improves on the $2$-factor. 

\begin{theorem}\label{thm:algo}
    There is a randomized polynomial-time algorithm for \FEC with approximation factor $1.78$ on simple graphs and $1+\sqrt{e}<1.607$ on simple bipartite graphs, where the approximation guarantees are in expectation and relative to a polynomial-time solvable LP. 
\end{theorem}

The main message of this work is the use of disjunctive programming techniques in approximation algorithms: exact local convexification via disjunctive programming techniques is strong enough to beat the naive $2$-factor for \FEC. We will elaborate on the LP and the rounding strategy shortly. Our approach leads to a new LP-relaxation for \pqFGC---we discuss this in Section \ref{sec:FGC}. We hope that the rounding algorithm for the disjunctive programming based LP for \FEC could serve as a stepping stone towards a constant factor approximation for \pqFGC via our new LP-relaxation. 

\paragraph{Discussion on Disjunctive Programming.} Disjunctive programming, developed by Balas in the 1970s, provides a systematic framework for exploiting disjunctive alternatives in integer programming. Its convexification principle yields extended formulations for the convex hulls of finite unions of polytopes \cite{B79}, providing a general tool for constructing strong LP relaxations. 
The 
convexification principle also underlies split cuts, an important family of cutting planes in integer programming. Split cuts are derived from the alternatives $\pi^Tx\le \pi_0$ or $\pi^Tx \ge \pi_0+1$ for integral vector $\pi$ and integer scalar $\pi_0$. The split closure intersects the convex hulls obtained from all such splits and is a rational polyhedron if the original polyhedron is rational \cite{CCK90}. Split cuts are close counterparts of Chv\'{a}tal-Gomory (CG) cuts, and have a rich theory comparing closure strength and rank (e.g., see \cite{CDDFG15, BCDJ21}). 

Disjunctive programming has found numerous applications in combinatorial optimization \cite{Let01,BZ04,Zuc04,CW08,A06, BZ18, FHW21, Bie08,BM12,Pri10,BSY13,FM18}. To the best of authors' knowledge, there are only a handful of applications of disjunctive programming in approximation algorithms. 
Bienstock \cite{Bie08} and Bienstock--McClosky \cite{BM12} used disjunctions to construct approximate extended LP formulations for knapsack and certain fixed-charge flow sets. Pritchard \cite{Pri10} extended this approach to fixed-dimensional knapsack. Bienstock, Sethuraman, and Ye \cite{BSY13} combined disjunctive formulations with rounding to obtain approximation algorithms for incremental knapsack. Faenza and Malinovi\'{c} \cite{FM18} rounded a disjunctive programming-based LP to design a PTAS for time-invariant incremental knapsack. 
Our work adds to this relatively short known list of applications of disjunctive programming in approximation algorithms.

\paragraph{Extensions.} Our techniques can be generalized to obtain a tighter $1.73$-approximation even in multigraphs. This involves strengthening the LP further using configuration variables. The strengthening complicates the presentation and we avoid these details in the interests of illustrating the main message of the work---namely, disjunctive programming is a concrete approach to design approximation algorithms for flexible network design. 



\subsection{Techniques and AI Disclosure}\label{sec:lp}
The hardness result is via reduction from vertex cover in cubic graphs. We describe the techniques for our approximation result henceforth. 
We first describe the natural LP-relaxation and the need to strengthen it. 
Let $G=(V, E=S\uplus U)$ be an undirected graph whose edges are partitioned into safe edges $S$ and unsafe edges $U$. 
We write $\delta(v)=\delta_E(v)$ and for a vector $z\in \R^E$ and an edge set $A$, let $z(A):=\sum_{e\in A}z_e$. Introducing an edge variable $x_e$ for each $e\in E$ to indicate whether the edge $e$ is chosen and a vertex variable $\lambda_v$ for each $v\in V$ to indicate whether the vertex $v$ is safe-covered gives the following LP-relaxation: 
    \begin{align*}\tag{\textit{LP}$_1$}\label{eq:LP1}
        \min \quad&\sum_{e\in E} c_e x_e\\
        \text{s.t.} \quad& x(\delta_{S}(v)) \geq p \lambda_{v} & \forall v \in V\\
        & x(\delta(v)) \geq (p+q) (1-\lambda_{v}) & \forall v \in V\\
        & 0 \leq  \lambda_{v} \leq 1 &\forall v \in V\\
        & 0 \leq x_{e} \leq 1 & \forall e \in E.
    \end{align*}
While \eqref{eq:LP1} is natural, its integrality gap is unbounded---see Section \ref{sec:lp1-integ-gap} for an unbounded integrality gap instance.

We use techniques from disjunctive programming to formulate a tighter LP-relaxation now.  For each $v \in V$, we define discrete sets 
    \begin{align*}
    \mathcal{A}_v &:= \{\alpha\in \{0,1\}^{\delta(v)}: \alpha(\delta_S(v))\ge p\},\\
    \mathcal{B}_v &:= \{\beta\in \{0,1\}^{\delta(v)}: \beta(\delta(v))\ge p+q\}. 
    \end{align*}
The convex hulls of both $\mathcal{A}_v$ and $\mathcal{B}_v$ are integral polytopes that admit a polyhedral description: namely the box constraints $[0,1]^{\delta(v)}$ and the lower bound constraint. Applying Balas' formulation to the union of these two convex hulls gives our LP-relaxation. Concretely, in addition to edge and vertex indicator variables as above, we introduce two edge-variable vectors $\alpha^v, \beta^v\in\{0,1\}^{\delta(v)}$ for each vertex $v\in V$, and consider the following LP-relaxation: 
\begin{align*}\tag{\textit{LP}$_2$}\label{eq:LP2}
        \min \quad&\sum_{e\in E} c_e x_e\\
        \text{s.t.}\quad& x_{e} = \alpha^{v}_{e} + \beta^{v}_{e} &\forall e \in \delta(v), v\in V \\
        & \alpha^{v}(\delta_{S}(v)) \geq \lambda_{v} p &\forall v \in V \\
        & \beta^{v}(\delta(v)) \geq (1-\lambda_{v})(p+q) &\forall v \in V \\
        & 0 \leq \alpha^{v}_{e} \leq \lambda_{v} &\forall e \in \delta(v), v\in V\\
        & 0 \leq \beta^{v}_{e} \leq 1 - \lambda_{v} &\forall e \in  \delta(v), v\in V\\
        & 0 \leq \lambda_{v} \leq 1 &\forall v \in V
\end{align*}
The formulation has $O(|V|+|E|)$ variables and constraints and is hence solvable in polynomial time. 
We first show that \eqref{eq:LP2} is indeed a relaxation of \FEC. 
\begin{proposition}\label{prop:relax}
    Let $F \subseteq E$ be a feasible \FEC on instance $(G=(V, E=S\uplus U), p, q)$. Then, 
    there exist $\lambda\in \{0,1\}^V, \alpha^v\in \{0,1\}^{\delta(v)}, \beta^v\in \{0,1\}^{\delta(v)}$ for each $v \in V$ such that $(\chi^F, \alpha, \beta, \lambda)$ is feasible to \eqref{eq:LP2}, where, $\chi^F\in \{0,1\}^E$ is the indicator vector of $F$. 
\end{proposition}
\begin{proof}
    For each $v\in V$, we set $\lambda_v=1$ if $|\delta_S(v)\cap F|\ge p$ and $\lambda_v=0$ otherwise. For each $v\in V, e\in \delta(v)$, we set 
    \begin{align*}
        \alpha^v_e &:=
        \begin{cases}
            1 &\text{ if } e\in F \text{ and } |\delta_S(v)\cap F|\ge p, \\
            0 &\text{ otherwise,}
        \end{cases}\\
        \beta^v_e &:=
        \begin{cases}
            1 &\text{ if } e\in F \text{ and } |\delta_S(v)\cap F|<p, \\
            0 &\text{ otherwise.}
        \end{cases}
    \end{align*}
    The resulting point $(\chi^F, \alpha, \beta, \lambda)$ satisfies the constraints of \eqref{eq:LP2}. 
\end{proof}

The projection of \eqref{eq:LP2} to $x$-variables is exactly the set $x$ of vectors whose restriction to $\delta(v)$ belongs to $\conv(\mathcal{A}_v \cup \mathcal{B}_v)$ for every $v\in V$. We formalize this now. 
For a vector $x\in \R^E$ and a vertex $v\in V$, we denote the restriction of the vector $x$ to the variables corresponding to $\delta(v)$ by $x_{\delta(v)}$. 

\begin{lemma}\label{lem:convex}
      Given a point $(x,\alpha,\beta, \lambda)$ that is feasible for \eqref{eq:LP2}, for each vertex $v \in V$, 
      \begin{enumerate}[label={(\arabic*)}]
          \item there exist $a^v \in \conv(\mathcal{A}_{v})$, $b^v \in \conv(\mathcal{B}_{v})$ such that $x_{\delta(v)} = \lambda_v a^v + (1-\lambda_v) b^v$ and 
          \item the vector $x_{\delta(v)}$ is a convex combination of vectors in $\calA_v\cup \calB_v$ and such a convex combination can be computed in polynomial time. 
      \end{enumerate}
\end{lemma}
\begin{proof}
    By the constraints, we have that $x_{\delta(v)}=\alpha^v+\beta^v$, where $\alpha^v\in \lambda_v\conv(\calA_v)$ and $\beta^v\in (1-\lambda_v)\conv(\calB_v)$. If $\lambda_v\in (0,1)$, then taking $a^v=\alpha^v/\lambda_v$ and $b^v = \beta^v/(1-\lambda_v)$ gives the needed conclusion. If $\lambda_v=0$, then take $a^v=0$ and $b^v=\beta^v$ and if $\lambda_v=1$, then take $a^v=\alpha^v$ and $b^v=0$ to obtain the desired conclusion. 
    Thus, we have $a^v \in \conv(\mathcal{A}_{v})$, $b^v \in \conv(\mathcal{B}_{v})$ such that $x_{\delta(v)} = \lambda_v a^v + (1-\lambda_v) b^v$. 
    We observe that $a^v, b^v$ are computable in polynomial time. 

    Next, we observe that $a^v\in \conv(\calA_v)$ and $\conv(\calA_v)$ is an integral polytope whose extreme points are vectors in $\calA_v$. Thus, by Caratheodary's theorem, $a^v$ is a convex combination of vectors in $\calA_v$ and such a convex combination can be computed in polynomial time (see \cite{schrijver-book-1}). Similarly, $b^v$ is a convex combination of vectors in $\calB_v$ and such a convex combination can be computed in polynomial time. Finally, we note that $x$ is a convex combination of $a^v$ and $b^v$ and thus, $x$ is a convex combination of vectors in $\calA_v \cup \calB_v$ and such a convex combination can be computed in polynomial time. 
\end{proof}

\paragraph{Rounding.} 
We interpret the decomposition of $x_{\delta(v)}$ in Lemma \ref{lem:convex}(2) as a local distribution over feasible local edge sets with edge marginals $x_e$. In the first phase, we choose a set $L$ of vertices suitably---the precise way to choose $L$ will be described shortly. Next, we independently sample a local cover at each vertex of $L$ according to the local distribution. Their union $(p,q)$-flexible covers every vertex in $L$ and let the union be denoted as $F$. In the second phase, we address the coverage of remaining vertices. For each remaining vertex, we add a cheapest set of incident edges to $F$ to achieve either safe coverage or cumulative coverage---this is easy to compute via greedy. The non-trivial issue is bounding the augmentation cost in the second phase by exploiting the coverage already supplied by the sampled local covers for vertices in $L$. We next explain how to pick the set $L$ and the analysis. 

We first focus on simple bipartite graphs since the choice of $L$ and the analysis are easier to describe. Our ideas for general graphs originated by considering bipartite graphs. Here, we choose $L$ to be one side of the bipartition. Then, the expected sampling cost in the first phase is exactly $\sum_{e\in E}c_e x_e$. At each vertex on the other side, distinct incident edges are supplied by distinct independently sampled neighbors owing to simplicity of the graph. This independence allows us to bound the expected augmentation cost in the second phase at each remaining vertex $u\in V\setminus L$ by at most $\frac{1}{\sqrt{e}}\sum_{e\in \delta(u)}c_e x_e$ (this is the non-trivial part of the argument). Thus, the total expected cost is $\left(1+\frac{1}{\sqrt{e}}\right)$ times the cost of the LP. 

For general simple graphs (that may not be bipartite), we include each vertex into $L$ independently with probability $\theta$. Then, the expected sampling cost is at most $2\theta \sum_{e\in E}c_e x_e$. Conditional on a vertex $u$ lying outside $L$, its incident edge $e$ is independently chosen in the first phase set $F$ with probability $\theta x_e$; we use this fact to show that the expected augmentation cost in the second phase at each remaining vertex $u\in V\setminus L$ is at most $e^{-\theta/2}\sum_{e\in \delta(u)}c_e x_e$. Thus, the total expected cost is $(2\theta + 2(1-\theta)e^{-\theta/2})\sum_{e\in E}c_e x_e$ leading to an approximation factor of 
\[
2\theta + 2(1-\theta)e^{-\theta/2}.
\]
Choosing $\theta \approx 0.47008$ makes this expression to be about $1.78$. Again, simplicity is used precisely to obtain independence among the contributions to an unsampled vertex; parallel edges can be supplied by the same neighbor leading to correlated inclusions and we avoid this via the simplicity assumption. 

\paragraph{AI Disclosure.}
ChatGPT 5.5 was used to prove the moment bound in Lemma \ref{lem:mom_bound} and it suggested reducing from minimum cardinality vertex cover in simple cubic graphs to obtain the hardness result stated in Theorem \ref{thm:hardness}. 
ChatGPT 6.1 Sol was used to obtain the unbounded integrality gap instance for \ref{eq:LP1} stated in Section \ref{sec:lp1-integ-gap} and ChatGPT Astra was used to assist with literature survey and the writing of the introduction. 

\paragraph{Related Work.}
Several variants of edge cover have been investigated and we mention a few. Vertex-specific degree requirements instead of uniform requirements as in \pEC remains tractable through $b$-matching \cite{schrijver-book}. In Colorful Edge Cover problem, vertices are colored and the goal is to cover at least $r_i$ vertices of color $i$ for each color $i$ using a minimum sized set of edges and this problem is also tractable \cite{BBB23}. Activation Edge Cover aims to minimize vertex payments to activate edges---it is NP-hard and has been studied from the perspective of approximation \cite{KNS22}. An especially relevant variant to our work is the demand edge cover problem introduced by Pritchard and Chakrabarty \cite{PC2011}. In this problem, each edge has a value and each vertex has a demand and the goal is to find a minimum cost subset of edges such that the sum of the values of edges incident to the vertex meets the demand of the vertex for all vertices. They prove that it is impossible to obtain a $(17/16-\epsilon)$-approximation for this problem for every constant $\epsilon>0$. It is possible to show APX-hardness of \FEC by reducing from demand edge cover but the resulting instances are multigraph instances and not simple. The reduction that we present shows APX-hardness for simple graphs. 

%% file: Hardness.tex
\section{Hardnesss and Inapproximability}
We show APX-hardness of \FEC on simple bipartite graphs in this section. 
We reduce \VC in cubic graphs to \FEC to prove Theorem \ref{thm:hardness}. 
In \VC, the input is a graph $G=(V, E)$ and the goal is to find a minimum cardinality subset $S$ of vertices such that every edge has at least one end-vertex in $S$. 
Chleb\'{i}k and Chleb\'{i}kov\'{a} \cite{CC2006} showed that \VC in simple cubic graphs is NP-hard to approximate within a factor of $\frac{100}{99}-\epsilon$ for every $\epsilon>0$. 

\begin{proof}[Proof of Theorem \ref{thm:hardness}]
We reduce minimum cardinality vertex cover in simple cubic graphs to \FEC. Let $G=(V,E)$ be an instance of minimum cardinality vertex cover. We set $p =1$ and $q=2$, construct the graph $G' = (V',E')$ and cost function $c': E'\rightarrow \Z_{\ge 0}$ as follows. The set $V'$ consists of $3$ copies $s_v, r_v, r'_v$ of each vertex $v\in V$ and $5$ vertices $x_e, a_e^1, a_e^2, b_e^1, b_e^2$ for each edge $e\in E$. 
The set $E'$ consists of 
safe edges $s_vr_v$ with $c'_{s_vr_v}=2$ for each $v\in V$, 
safe edges $r_v r'_v$ with $c'_{r_v r'_v}=0$ for each $v\in V$, 
unsafe edges $x_e s_v$ and $x_e s_u$ with $c'_{x_e s_v}=c'_{x_e s_u}=1$ for each $e=uv\in E$, unsafe edges $x_e a_e^i$ with $c'_{x_e a_e^i}=0$ for each $i\in [2]$ and $e\in E$, safe edges $a_e^i b_e^i$ with $c'_{a_e^i b_e^i}=0$ for each $i\in [2]$ and $e\in E$. We note that the constructed graph $G'$ is simple bipartite with bipartition $X = \left\{x_{e}, b^{i}_{e}\right\}_{e \in E, i \in [2]} \cup \left\{r_{v}\right\}_{v \in V}$ and $V' \setminus X$. See Figure \ref{fig:fec-eight-partition} for an illustration of $G'$. 
Let $\tau(G)$ be the value of the smallest vertex cover in $G$ and $OPT$ be the optimal value of \FEC on $G'$. 

\input{fec-eight-partition-figure}


Let $C \subseteq V$ be a vertex cover of $G$. We construct a subset $F \subseteq E'$ as follows. For every vertex $v \in C$, add the three unsafe edges of the form $x_{e} s_{v}$ to $F$. For every vertex $v \notin C$ add the safe edges $s_{v} r_{v}$ and $r_{v} r'_{v}$ to $F$. Lastly add the edges $x_{e} a^{i}_e$ and $a^{i}_e b^{i}_{e}$ for all $e \in E$, $i \in \left\{1,2\right\}$. We observe that $F$ is feasible for \FEC on instance $(G', p, q, c')$. We have that $c'(F) = 3|C| + 2(|V|-|C|) = 2|V| + |C|$, and hence, $OPT \leq 2|V| + \tau(G)$. 

Next, let $F$ be a feasible solution to the \FEC instance $(G', p, q, c')$. Let $C_{F} = \left\{v \in V \mid s_{v} x_{e} \in F \text{ for some } e \in E\right\}$. We show that $C_{F}$ is a vertex cover in $G$. Fix an edge $e \in E$. Since the vertex $x_{e}$ is incident to $4$ edges, which are all unsafe, we must have that $d_{F}(x_{e}) \geq 3$ by definition of $F$. Since $2$ of the edges incident to $x_{e}$ are of the form $s_{v} x_{e}$, by the pigeonhole principle, $F$ contains an edge $s_{u} x_{e}$ for some vertex $u \in V$. Since $u$ covers $e$ and  $u \in C_{F}$, $C_{F}$ is a vertex cover. Next, we bound the size of the vertex cover $C_F$. If $v \in V \setminus C_{F}$, then edge $s_{v} r_{v}$ of cost $2$ is in $F$ and so $c'(\delta_{F}(s_v)) \geq 2$. Otherwise, suppose $v \in C_{F}$. This implies that $s_{v} x_{e'} \in \delta_{F}(s_v)$ for some $e' \in E$. If $s_{v} r_{v} \in \delta_{F}(s_v)$, then $c'(\delta_{F}(s_v)) \geq 3$. Otherwise $s_{v} r_{v} \notin \delta_{F}(s_v)$ and so $\delta_{F}(s_v)$ must contain the three edges of the form $s_{v} x_e$, which again implies $c'(\delta_{F}(s_v)) \geq 3$. Thus, we conclude that
\begin{equation}\label{eq:inequality}
    c'(F) \geq \sum_{v \in V} c'(\delta_{F}(v)) \geq 3|C_{F}| + 2|V \setminus C_{F}| = 2|V| + |C_{F}|
\end{equation}
Therefore, we have that OPT $\geq 2|V| + \tau(G)$. Since we showed earlier that the reverse inequality holds, we have that
 \begin{equation}\label{eq:equality}
     OPT = 2|V| + \tau(G) 
 \end{equation}
        
For all simple graphs $H = (V_H,E_H)$, we have that $\tau(H) \geq \frac{|E_H|} {\Delta(H)}$. This follows since given any vertex cover $C$ of $H$, we have that $|C| (\Delta(H)) \geq \sum_{v \in C} d_{H}(v) \geq |E_H|$. If $H$ is a cubic graph we have that $\tau(H) \geq \frac{|V_H|}{2}$. Thus, $2|V_H|  + \tau(H) \leq 5 \tau (H)$ for all cubic graphs $H$. Suppose we had a $\rho$-approximation to \FEC on instance $(G', p, q, c')$ that returned a solution $F$. From the previous analysis, $C_{F}$ is a vertex cover in $G$. We have that 
\begin{align}
    |C_{F}| &\leq c'(F) - 2|V| \label{bound_2}\\
    &\leq \rho OPT - 2|V|\notag \\
    &=  \rho (2|V|  +\tau(G)) - 2|V| \label{bound_3}\\
    &= (\rho - 1)(2|V| + \tau(G)) + \tau(G)\notag\\
    &\leq (\rho - 1)(5 \tau(G)) + \tau(G)\label{bound_4}\\
    & = (5 \rho - 4) \tau(G)\notag
\end{align}
where \eqref{bound_2} follows from \eqref{eq:inequality}, \eqref{bound_3} follows from \eqref{eq:equality}, and \eqref{bound_4} follows from the fact that $G$ is cubic. Thus, a $\rho$-approximation for \FEC would give a $5 \rho - 4$ approximation to minimum cardinality vertex cover on cubic graphs. This implies that a $\frac{496}{495} - \epsilon$ approximation for \FEC would yield a $5\left(\frac{496}{495} - \epsilon\right)-4 = \frac{100}{99} - 5 \epsilon$ approximation for \VC, which is NP-hard via the results of Chleb\'{i}k and Chleb\'{i}kov\'{a} \cite{CC2006}.
\end{proof}

%% file: fec-eight-partition-figure.tex
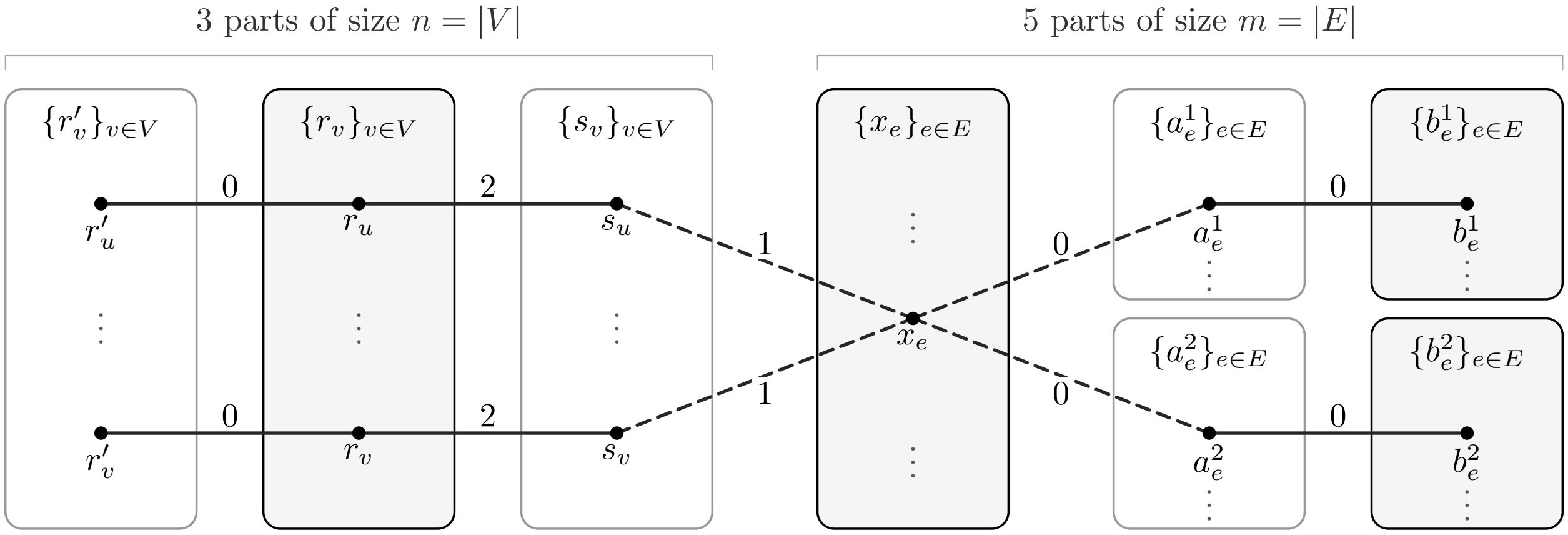
\begin{figure*}[t]
\centering
\resizebox{\linewidth}{!}{%
\begin{tikzpicture}[
  font=\small,
  part/.style={rounded corners=5pt,line width=.65pt},
  partX/.style={part,draw=black!50!black,fill=black!4},
  partY/.style={part,draw=white!60!black,fill=white!8},
  family/.style={font=\small,text=black},
  vertex/.style={circle,fill=black,draw=black,inner sep=0pt,
                 minimum size=3.6pt},
  vlabel/.style={font=\small,inner sep=1.2pt,text=black},
  safe/.style={draw=black!85,line width=1pt,line cap=round},
  unsafe/.style={draw=black!85,line width=1pt,line cap=round,
                dash pattern=on 4pt off 2.5pt},
  cost/.style={font=\small,fill=white,inner sep=1.5pt},
  group/.style={font=\small,text=black!80},
  omitted/.style={font=\small,text=black!65}
]

\draw[partY] (-1,-2.45) rectangle (1,2.15);
\draw[partX] (1.7,-2.45) rectangle (3.7,2.15);
\draw[partY] (4.4,-2.45) rectangle (6.4,2.15);
\draw[partX] (7.5,-2.45) rectangle (9.5,2.15);
\draw[partY] (10.6,-.05) rectangle (12.6,2.15);
\draw[partY] (10.6,-2.45) rectangle (12.6,-.25);
\draw[partX] (13.3,-.05) rectangle (15.3,2.15);
\draw[partX] (13.3,-2.45) rectangle (15.3,-.25);

\node[family] at (0,1.8) {$\{r'_v\}_{v\in V}$};
\node[family] at (2.7,1.8) {$\{r_v\}_{v\in V}$};
\node[family] at (5.4,1.8) {$\{s_v\}_{v\in V}$};
\node[family] at (8.5,1.8) {$\{x_e\}_{e\in E}$};
\node[family] at (11.6,1.8) {$\{a_e^1\}_{e\in E}$};
\node[family] at (11.6,-.6) {$\{a_e^2\}_{e\in E}$};
\node[family] at (14.3,1.8) {$\{b_e^1\}_{e\in E}$};
\node[family] at (14.3,-.6) {$\{b_e^2\}_{e\in E}$};

\draw[black!35,line width=.4pt]
  (-1,2.35) -- (-1,2.5) -- (6.4,2.5) -- (6.4,2.35);
\draw[black!35,line width=.4pt]
  (7.5,2.35) -- (7.5,2.5) -- (15.3,2.5) -- (15.3,2.35);
\node[group] at (2.7,2.85) {3 parts of size $n=|V|$};
\node[group] at (11.4,2.85) {5 parts of size $m=|E|$};

\coordinate (rpu) at (0,.95);
\coordinate (rpv) at (0,-1.45);
\coordinate (ru) at (2.7,.95);
\coordinate (rv) at (2.7,-1.45);
\coordinate (su) at (5.4,.95);
\coordinate (sv) at (5.4,-1.45);
\coordinate (xe) at (8.5,-.25);
\coordinate (a1) at (11.6,.95);
\coordinate (a2) at (11.6,-1.45);
\coordinate (b1) at (14.3,.95);
\coordinate (b2) at (14.3,-1.45);

\draw[safe] (rpu) -- node[cost,above] {0} (ru);
\draw[safe] (rpv) -- node[cost,above] {0} (rv);
\draw[safe] (ru) -- node[cost,above] {2} (su);
\draw[safe] (rv) -- node[cost,above] {2} (sv);
\draw[unsafe] (su) -- node[cost,above] {1} (xe);
\draw[unsafe] (sv) -- node[cost,below] {1} (xe);
\draw[unsafe] (xe) -- node[cost,above] {0} (a1);
\draw[unsafe] (xe) -- node[cost,below] {0} (a2);
\draw[safe] (a1) -- node[cost,above] {0} (b1);
\draw[safe] (a2) -- node[cost,above] {0} (b2);

\foreach \id/\lab in {rpu/{r'_u},rpv/{r'_v},ru/{r_u},rv/{r_v},
                       su/{s_u},sv/{s_v},xe/{x_e},
                       a1/{a_e^1},a2/{a_e^2},b1/{b_e^1},b2/{b_e^2}} {
  \node[vertex,label={[vlabel]below:$\lab$}] at (\id) {};
}
\foreach \xx in {0,2.7,5.4}
  \node[omitted] at (\xx,-.25) {$\vdots$};
\node[omitted] at (8.5,.8) {$\vdots$};
\node[omitted] at (8.5,-1.65) {$\vdots$};
\foreach \xx in {11.6,14.3} {
  \node[omitted] at (\xx,.3) {$\vdots$};
  \node[omitted] at (\xx,-2.1) {$\vdots$};
}

\end{tikzpicture}%
}
\caption{
An illustration of the construction of $G'$. Continuous edges are safe and dashed edges are unsafe. Edge labels refer to the costs. The shaded parts and the unshaded parts show a bipartition of $G'$. }
\label{fig:fec-eight-partition}
\end{figure*}

%% file: Algorithms-2.tex
\section{Approximation Algorithms}\label{sec:algo}
In this section, we design approximation algorithms for \FEC in simple graphs to prove Theorem \ref{thm:algo}. 
Our algorithm starts with an optimum solution $(x, \alpha, \beta, \lambda)$ to \eqref{eq:LP2} and will round it. The rounding algorithm additionally takes a subset $L\subseteq V$ as input whose significance will be clear shortly. 
The rounding algorithm proceeds in two phases. 
The algorithm addresses the covering requirements of vertices in $L$ in the first phase as follows: let $v\in L$. By Lemma \ref{lem:convex}, $x_{\delta(v)}$ is a convex combination of safe-satisfying vectors and cumulative-satisfying vectors (i.e., vectors in $\calA_v$ and $\calB_v$). The algorithm samples a vector $F_v$ among all vectors in the convex combination with probability proportional to its coefficient in the convex combination. It sets the solution of the first phase as $F_1=\cup_{v\in L}F_v$. We observe that each $v\in L$ is either safe-covered or cumulative-covered by $F_v$ and hence, by $F_1$ and moreover, the expected cost of $F_1$ is at most the cost of the LP solution (since $L$ is an independent set and consequently $\delta(a)\cap \delta(b)=\emptyset$ for every $a, b\in L$). However, we could have vertices $u\in V\setminus L$ that are uncovered by $F_1$ (i.e., neither cumulative-covered nor safe-covered). The algorithm addresses the covering-requirements for vertices in $V\setminus L$ in the second phase as follows: for each $u\in V\setminus L$, it computes a cheapest augmentation set $A_u^1$ to $F_1$ that makes $u$ safe-covered (by greedy), a cheapest augmentation set $A_u^2$ to $F_1$ that makes $u$ cumulative-covered (by greedy), and sets $A_u:=\arg\min\{c(A_u^1), c(A_u^2)\}$. It sets the solution of the second phase as $F_2=\cup_{u\in R}A_u$ and returns $F_1\cup F_2$. See Algorithm \ref{algo:round} for a formal description. 

    \begin{algorithm}[h]
    \caption{Rounding Algorithm}
    \label{algo:round}
        \begin{algorithmic}[1]
            \State \textbf{Input:} $G = (V,E = S \sqcup U), c: E \rightarrow \mathbb{R}_{\geq 0}, p, q$, Feasible solution $(x, \alpha, \beta, \lambda)$ to \eqref{eq:LP2}, a subset $L\subseteq V$. 
            \For {each $v \in L$}
                \State Express $x_{\delta(v)}$ as a convex combination of vectors in $\calA_v\cup \calB_v$, \newline \indent say $x_{\delta(v)}=\sum_{a\in \calA_v \cup \calB_v}\lambda_a a$ with $\sum_{a\in \calA_v\cup \calB_v}\lambda_a=1$ and $\lambda_a\ge 0$ for all $a\in \calA_v\cup \calB_v$ 
                \State Sample $a\in \calA_v\cup\calB_v$ with probability proportional to $\lambda_a$ and set $F_v=\text{support}(a)$
            \EndFor
            \State $F_1 \leftarrow \cup_{v\in L} F_{v}$
            \For {each $u \in V\setminus L$}
                \State $A_{u}^{1}\leftarrow \arg\min \left\{c(A) : A \subseteq \delta_{S}(u) \setminus F_1, \,\ |(\delta_{(F_1\cup A)\cap S}(u)) | \geq p\right\}$
                \State $A_{u}^{2} \leftarrow \arg\min \left\{c(A) : A \subseteq \delta(u) \setminus F_1, \,\ |\delta_{F_1\cup A}(u) | \geq p+q\right\}$
                \State $A_u\leftarrow \arg\min\{c(A_u^1), c(A_u^2)\}$
            \EndFor
            \State $F_2\leftarrow \cup_{u\in V\setminus L}A_u$
            \State \Return $F:=F_1 \cup F_2$
        \end{algorithmic}
    \end{algorithm}

Essentially, the algorithm uses the LP solution to sample and satisfy the covering requirements for vertices in $L$ in the first phase and in the second phase, it augments the solution from the first phase with additional edges to meet the covering requirements for each vertex in $V\setminus L$. 
If $L=\emptyset$, then the algorithm does not take advantage of the LP solution and simply corresponds to the naive greedy algorithm mentioned in the introduction which yields a $2$-approximation which is also the best possible. 
Our key contribution is an LP-informed greedy augmentation that improves on the $2$-factor. 
Our choice for $L$ varies between bipartite graphs and non-bipartite graphs and we will discuss it shortly. 

We first show that Algorithm \ref{algo:round} can be implemented to run in polynomial time and argue its correctness. 
\begin{lemma}\label{algo:poly-time}
Algorithm \ref{algo:round} can be implemented to run in polynomial time. 
\end{lemma}
\begin{proof}
Step 3 can be implemented in polynomial time via Lemma \ref{lem:convex}; we note that the convex combination has at most polynomially many non-zero coefficients. Consequently, the sampling procedure in Step 4 is also implementable in polynomial time. 
Steps 8 and 9 are greedy procedures that can be implemented in polynomial time via a sorting algorithm. Thus, the entire algorithm can be implemented in polynomial time. Next, we show that the algorithm returns a feasible solution. 
\end{proof}
We now prove correctness of the algorithm. 
\begin{lemma}\label{lem:algo-correctness}
For every choice of $L\subseteq V$, Algorithm \ref{algo:round} returns a feasible flexible-edge-cover for the instance $(G, p, q)$. 
\end{lemma}
\begin{proof}
    Let $F$ be the solution returned by the algorithm for the input instance $(G, p, q)$ with a feasible solution $(x, \alpha, \beta, \lambda)$ to \eqref{eq:LP2}. We show that every vertex $v\in L\cup R$ is either safe-covered or cumulative-covered by $F$. 
    
    Let $v \in L$. We recall that $x_{\delta(v)}=\sum_{a\in \calA_v\cup \calB_v}a\lambda_a$ with $\sum_{a\in \calA_v\cup \calB_v}\lambda_a=1$ and $\lambda_a\ge 0$ for all $a\in \calA_v\cup\calB_v$. For each $a\in \calA_v$, we have that $v$ is safe-covered by support$(a)$ and for each $a\in \calB_v$, we have that $v$ is cumulative-covered by support$(a)$. Thus, for each $a\in \calA_v\cup\calB_v$, we have that $v$ is either safe-covered or cumulative-covered by support$(a)$. Hence, $v$ is either safe-covered or sumulative-covered by $F_1\subseteq F$. 
    
    Next, consider $u\in R$. We have two cases: 
        (i) Suppose $A_u=A_u^1$. We recall that 
        \[A^{1}_{u} = \arg\min \left\{c(A) : A \subseteq \delta_{S}(u) \setminus F_1, \,\ |(\delta_{(F_1\cap S)\cup A}(u)) | \geq p\right\}\] and hence, $|\delta_{(F_1\cup A_u^1)\cap S}(u)|\ge p$. Consequently, $u$ is safe-satisfied by $F_1\cup A_u\subseteq F$. 
        (ii) Suppose $A_u=A_u^2$. We recall that \[A^{2}_{u} = \arg\min \left\{c(A) : A \subseteq \delta(u) \setminus F_1, \,\ |(\delta_{F_1\cup A}(u)) | \geq p+q\right\}\] and hence, $|\delta_{F_1\cup A_u^2}(u)|\ge p+q$. Consequently, $u$ is cumulative-satisfied by $F_1\cup A_u\subseteq F$.    
\end{proof}

The rest of the section is devoted to analyzing the approximation factor via a careful choice of $L$ in bipartite and non-bipartite graphs. 
All graphs in the remainder of this section are simple. 
To analyze the approximation factor, we address the expected cost of the solution from the two phases. In the first phase, our sampling strategy depends on the LP solution which allows us to show that the cost of $F_1$ is small relative to the LP. In the second step, we augment $F_1$ with additional edges in a greedy manner to fix the covering requirements of vertices in $V\setminus L$. This augmentation step incurs additional cost that is non-trivial to bound. In Section \ref{sec:key-technical-lemma}, we prove a key technical lemma that helps bound the cost incurred in an augmentation step in a general setting. In Section \ref{sec:augmentation-cost}, we adapt this technical lemma to the context of \FEC to bound the cost incurred by the augmentation step relative to the objective value of \eqref{eq:LP2}. 
In Section \ref{sec:alg-bipartite}, we specify the choice of $L$ and design a $1.606$-approximation for simple bipartite graphs. 
In Section \ref{sec:alg-arbitrary}, we specify the choice of $L$ and design a $1.78$-approximation for simple non-bipartite graphs. 
The rounding strategy for bipartite graphs is slightly simpler and its analysis served as the starting point of our rounding strategy for non-bipartite graphs, so we encourage readers to read Section \ref{sec:alg-bipartite} on bipartite graphs as a precursor to reading Section \ref{sec:alg-arbitrary} for non-bipartite graphs. For a real value $a$, we use the convention that $(a)^+=\max\{a, 0\}$. 
\input{technical-lemma}

\input{bipartite}
\input{non-bipartite}

%% file: technical-lemma.tex
\subsection{Key Technical Lemma}\label{sec:key-technical-lemma}
In this section, we show the following key technical lemma that compares the cost of the cheapest $r$-element subset to the cost of the cheapest augmentation $A$ to a randomly sampled subset $F$ to achieve $|A\cup F|\ge r$. 
    \begin{lemma}\label{lem:key}
     Let $N$ be a ground set with cost function $c: N\rightarrow \R_{\ge 0}$ and $r\in \Z_+$ be a positive integer. 
     Let $C_r=\min\{c(C): C\subseteq N, |C|=r\}$, i.e., $C_r$ is the sum of the costs of the cheapest $r$ elements of $N$. 
     Let $\alpha>0$ and $x,y\in [0,1]^N$ such that $0\le \alpha y_i\le x_i\le 1$ for all $i\in N$ and $y(N)\ge r$. Let $F\subseteq N$ be a random set obtained by selecting each element $i\in N$ with probability $x_i$ independently. 
     Then, 
     \[
     E_{F}\left[\min\left\{c(F') : F' \subseteq N \setminus F,\, |F' \cup F| \geq r\right\}\right] \le e^{-\alpha}C_r.
     \]
     \end{lemma}

We need the following helper lemma to prove Lemma \ref{lem:key}.
    \begin{lemma}\label{lem:mom_bound}
     Let $\left\{X_i\right\}_{i=1}^{n}$ be independent Bernoulli variables with $x_i=E[X_i]$, $\mu=\sum_{i=1}^n x_i$, and $Y:=\sum_{i=1}^n X_i$. For every non-negative integer $a\in \Z_{> 0}$, we have that 
     \[
        E[(a-Y)^+]\le ae^{-\frac{\mu}{a}}.
     \]
    \end{lemma}
    \begin{proof}
        Let $y, a \in \mathbb{Z}_{\geq 0}$. We have that $a(1- \frac{y}{a}) \leq a\left(1- \frac{1}{a}\right)^{y}$ by convexity of the function $f:[0,1] \rightarrow \mathbb{R}$ given by $f(x) = (1-x)^{y}$. Thus, we have that $(a-y)^{+} \leq a\left(1-\frac{1}{a}\right)^{y}$. Hence,
        \begin{align}
            E[(a-Y)^{+}] &\leq aE\left[\left(1-\frac{1}{a}\right)^{Y}\right]\notag\\
            &= a E \left[\left(1-\frac{1}{a}\right)^{\sum_{i=1}^{m} X_i}\right]\notag\\
            &= a \prod_{i=1}^{m} E\left[\left(1-\frac{1}{a}\right)^{X_i}\right]\label{eq:independence}\\
            &= a \prod_{i=1}^{m} \left(1 - \frac{x_i}{a}\right)\label{eq:binary}\\
            &\leq ae^{-\mu/a}.\label{ineq:convexity}
        \end{align}
     where \eqref{eq:independence} is by independence of $\left\{X_i\right\}_{i=1}^{n}$, \eqref{eq:binary} is because $X_i$ takes values in $\{0,1\}$ for every $i\in [n]$, and \eqref{ineq:convexity} is by convexity of the function $f(x)=e^x$.     
    \end{proof}

We now prove Lemma \ref{lem:key}. 
    \begin{proof}[Proof of Lemma \ref{lem:key}]
        
    For a real value $t$, we define $N_{\geq t} := \left\{i \in N : c_i \geq t\right\}$ and $N_{< t} := \left\{i \in N : c_i < t\right\}$. For a positive integer $r\in \Z_+$ and a real value $t$, we define $r_{t} = (r - |N_{< t}|)^{+}$. For a subset $F\subseteq N$ and a positive integer $r\in \Z_+$, we define $A_{F,r}:=\min\left\{c(F') : F' \subseteq N \setminus F,\, |F' \cup F| \geq r\right\}$. We define $Z_{F} = \text{argmin} \left\{c(F') : F' \subseteq N \setminus F,\, |F' \cup F| \geq r\right\}$. For a set $A \subset \R$, we denote the real-valued function $\mathbf{1}_{A}: \R \rightarrow \R$ as $\mathbf{1}_{A}(t) = 1$ if $t \in A$ and $\mathbf{1}_{A}(t) = 0$ if $t \notin A$.

         We need to show that $E_{F}[\sum_{i \in Z_F}c_i] \leq e^{-\alpha}C_r$. 
        Let us fix a set $F$ and consider $Z_F$. We can express $\sum_{i \in Z_F} c_i$ as $\sum_{i \in Z_F}\left(\int_{0}^{\infty} \mathbf{1}_{[0,c_i]}(t) \, dt\right)$. Swapping the order of the finite sum and the integral 
        gives 
        \[
        \sum_{i\in \Z_F} c_i = \int_{0}^{\infty}\sum_{i \in Z_F} \mathbf{1}_{[0,c_i]}(t) \, dt. 
        \]
        We will use the following claim to simplify $\sum_{i \in Z_F} \mathbf{1}_{[0,c_i]}$ for a fixed $t\ge 0$. 
\begin{claim}\label{claim:step_func}
        Let $t \geq 0$. Then, $\sum_{i \in Z_F} \mathbf{1}_{[0,c_i]}(t) = (r - |N_{<t}| - |F \cap N_{\ge t}|)^{+}$.
    \end{claim}
    \begin{proof}
        We note that $\sum_{i \in Z_{F}} \mathbf{1}_{[0,c_i]}(t)$ counts exactly the elements in $Z_{F}$ that have cost at least $t$. Since $Z_{F} \subseteq N \setminus F$, $\sum_{i \in Z_F}\mathbf{1}_{[0,c_i]}(t) = |Z_F \cap (N_{\geq t} \setminus F)|$. We first show that $|Z_{F} \cap (N_{\geq t} \setminus F)| = (|Z_{F}| - |(N_{<t} \setminus F)|)^{+}$ by considering two cases:
        
        Case (i). $|Z_{F}| > |N_{<t} \setminus F|$. Since $Z_{F}$ is a subset of $N \setminus F$ that contains the cheapest $|Z_{F}|$ items of $N \setminus F$, $Z_{F}$ contains all the the items in $N_{<t} \setminus F$ and hence contains the cheapest $|Z_{F} \setminus N_{<t}|$ items of $N_{\geq t} \setminus F$. Thus, in this case there are exactly $|Z_{F}| - |N_{<t}\setminus F|$ elements in $Z_{F}$ that have cost at least $t$.

        Case (ii). $|Z_{F}| \leq |N_{<t} \setminus F|$. In this case, $Z_{F}$ does not contain elements with cost at least $t$. Thus $Z_{F}$ has exactly $(|Z_{F}| - |(N_{<t} \setminus F)|)^{+}$ with cost at least $t$.

        Now, we have that
        \begin{align}
            (|Z_{F}| - |(N_{<t} \setminus F)|)^{+} &= (Z_{F} - |N_{<t}| + |F \cap N_{<t}|)^{+}\notag\\
            &= (r-|F|- |N_{<t}| + |F \cap N_{<t}|)^{+} \label{eq:cardinality}\\
            &= (r-|N_{<t}|-|F \cap N_{\geq t}|)^{+}\notag
        \end{align}
        where \eqref{eq:cardinality} follows from the fact that $|F \sqcup Z_{F}| = r$. 
    \end{proof}

    By Claim \ref{claim:step_func}, we have that 
    $$\sum_{i \in Z_F} c_i = \int_{0}^{\infty} \left(\sum_{i \in Z_F} \mathbf{1}_{[0,c_i]}(t)\right)\, dt = \int_{0}^{\infty} (r - |N_{<t}|-|F \cap N_{\geq t}|)^{+} \, dt.$$
    We note that $(r-|N_{<t}|-|F \cap N_{\geq t}|)^{+} \leq ((r- |N_{<t}|)^{+} - |F \cap N_{\geq t}|)^{+} = (r_t - |F \cap N_{\geq t}|)^{+}$, which implies that $$\sum_{i \in Z_F} c_i \leq \int_{0}^{\infty} (r_t - |F \cap N_{\geq t}|)^{+}\, dt.$$ 
Hence, we have that 
        $$E_F\left[\sum_{i\in Z_F}c_i\right]
        \le E_F\left[\int_{0}^{\infty} (r_t - |F \cap N_{\geq t}|)^{+}dt\right] = \int_{0}^{\infty} E_F\left[ (r_t - |F \cap N_{\geq t}|)^{+}\right]\, dt.$$
        Let us fix $t\ge 0$ and consider $E_F\left[ (r_t - |F \cap N_{\geq t}|)^{+}\right]$. Lemma \ref{lem:mom_bound} implies that for $r_t > 0$, $E\left[ (r_t - |F \cap N_{\geq t}|)^{+}\right] \leq r_t e^{-x(N_{\geq t})/r_t}$. If $r_{t} = 0$, then trivially $E\left[ (r_t - |F \cap N_{\geq t}|)^{+}\right] = 0$. Thus, we have that
        \begin{align}
            \int_{0}^{\infty} E\left[ (r_t - |F \cap N_{\geq t}|)^{+}\right]\, dt &\leq \int_{\left\{t : t \geq 0,\,r_t > 0\right\}} r_t e^{-x(N_{\geq t})/r_t} \, dt \notag\\
            &\leq  \int_{\left\{t : t \geq 0,\,r_t > 0\right\}} r_{t} e^{-\alpha y(N_{\geq t})/r_t}\, dt \label{ineq:y_hypo}\\
            &=  \int_{\left\{t : t \geq 0,\,r_t > 0\right\}} r_{t} e^{-\alpha (y(N) - y(N_{<t}))/r_t}\, dt \notag\\
            &\leq e^{-\alpha} \int_{\left\{t : t \geq 0,\,r_t > 0\right\}} r_t \, dt \label{ineq:r_pos}
        \end{align}
        where \eqref{ineq:y_hypo} follows from the hypothesis that $x_i \geq \alpha y_i$ for all $i \in N$ and \eqref{ineq:r_pos} follows from the fact that $y(N) - Y(N
        _{<t}) \geq r - |N_{<t}| = r_t$ for $r_t > 0$.       
        We now bound $\int_{0}^{\infty} r_{t}\, dt$. 
        Without loss of generality, let the items be ordered such that $c_1 \leq c_2 \cdots \leq c_{|N|}$ and let $c_0 = 0$. For $t \in (c_i,c_{i+1}]$, we have that $r_t = (r-N_{< t})^{+} = (r - i)^{+}$. Thus,
        \begin{align*}
            \int_{0}^{\infty} r_{t} \, dt &= \sum_{i=0}^{n-1} (r-i)^{+} (c_{i+1} - c_{i})\\
            &= \sum_{i=0}^{r-1} (r-i)(c_{i+1} - c_{i})\\
            &= \left(r c_{r} - (r-1)c_{r} + \sum_{i=1}^{r-1} c_i\right)\\
            &= C_{r}.
        \end{align*}
        Thus, $E_F[\sum_{i\in Z_F}c_i]\le e^{-\alpha}C_r$. 
    \end{proof}

    
\subsection{Augmentation Cost}\label{sec:augmentation-cost}
We now use Lemma \ref{lem:key} to bound the augmentation cost of covering a vertex $u$ after sampling the edges incident to it according to a distribution that depends on $x_{\delta(u)}$. 
\begin{lemma}\label{lem:augmentation-cost}
    Let $(G=(V, E), p, q, c: E\rightarrow \R_{\ge 0})$ be the input instance of \FEC, $(x, \alpha, \beta, \lambda)$ be a feasible solution to \eqref{eq:LP2}. Let $u\in V$ and $\theta\in [0, 1]$. Let $F_u\subseteq \delta(u)$ denote the set of edges obtained by sampling each edge $e\in \delta(u)$ independently with probability $\theta x_e$. 
    Let 
    \begin{align*}
    A_{u}^{1}&:=\arg\min \left\{c(A) : A \subseteq \delta_{S}(u) \setminus F_u, \,\ |(\delta_{(F_u\cup A)\cap S}(u)) | \geq p\right\}\\
    A_{u}^{2}&:=\arg\min \left\{c(A) : A \subseteq \delta(u) \setminus F_u, \,\ |\delta_{F_u\cup A}(u) | \geq p+q\right\}\\
    A_u&:=\arg\min\{c(A_u^1), c(A_u^2)\}.
    \end{align*}
    Then, 
    \[
    E[c(A_u)] \le e^{-\frac{\theta}{2}}\sum_{e\in \delta(u)}c_e x_e.
    \]
\end{lemma}
\begin{proof}[Proof of Lemma \ref{lem:augmentation-cost}]
We define
\begin{align*}
    C_p&:=\min\left\{c(F): F\subseteq \delta_S(u), |F|=p\right\}\\
    D_{p+q}&:=\min\left\{c(F): F\subseteq \delta(u), |F|=p+q\right\}. 
\end{align*}
By the first part of Lemma \ref{lem:convex}, we have that $x_{\delta(u)}=\lambda_u a^u + (1-\lambda_u)b^u$ where $a^u\in \conv(\calA_u)$ and $b^u\in \conv(\calB_u)$. Thus, $0\le \theta \lambda_u a^u\le \theta x_{\delta(u)}$ and $0\le \theta (1-\lambda_u)b^u\le x_{\delta(u)}\le 1$ and moreover, $a^u(\delta_S(u))\ge p$ and $b^u(\delta(u))\ge p+q$. 
Applying Lemma \ref{lem:key} to $N=\delta(u)$ with cost function $c:\delta(u)\rightarrow \R_{\ge 0}$, $r=p$, $\alpha=\lambda_u$, $y=a^u$, we obtain that 
\[
E[c(A_u^1)]\le e^{-\theta\lambda_u} C_p.
\]
Applying Lemma \ref{lem:key} to $N=\delta(u)$ with cost function $c:\delta(u)\rightarrow \R_{\ge 0}$, $r=p+q$, $\alpha=1-\lambda_u$, $y=b^u$, we obtain that 
\[
E[c(A_u^2)]\le e^{-\theta(1-\lambda_u)} D_{p+q}.
\]
Consequently, we have that 
\begin{align}
    E[c(A_u)]
    &=E\left[\min\left\{c(A_u^1), c(A_u^2)\right\}\right] \notag\\
    &\le \min \left\{E\left[c(A_u^1)\right], E\left[c(A_u^2)\right]\right\} \notag\\
    &\le \min \left\{e^{-\theta\lambda_u} C_p, e^{-\theta(1-\lambda_u)} D_{p+q}\right\} \notag\\
    &\le \left(e^{-\theta\lambda_u} C_p\right)^{\lambda_u}\left(e^{-\theta(1-\lambda_u)} D_{p+q}\right)^{1-\lambda_u} \label{ineq:min-atmost-gm}\\
    &= e^{-\theta(\lambda_u^2+(1-\lambda_u)^2)} C_p^{\lambda_u} D_{p+q}^{1-\lambda_u} \notag\\
    &\le e^{-\theta(\lambda_u^2+(1-\lambda_u)^2)} \left(\lambda_u C_p + (1-\lambda_u)D_{p+q}\right) \label{ineq:gm-atmost-am}\\
    &\le e^{-\frac{\theta}{2}}\left(\lambda_u C_p + (1-\lambda_u)D_{p+q}\right) \label{ineq:sqrt-e-bound}\\
    &\le e^{-\frac{\theta}{2}}\left({\lambda_u\sum_{e\in \delta(u)} c_e a^u_e+(1-\lambda_u)\sum_{e\in \delta(u)}c_eb^u_e}\right) \label{ineq:cheapest-to-lp}\\
    &=e^{-\frac{\theta}{2}}\left({\sum_{e\in \delta(u)}c_e x_e}\right), \label{ineq:x-bound}
\end{align}
where inequality \eqref{ineq:min-atmost-gm} is because the minimum of a collection of non-negative numbers is at most the geometric mean, inequality \eqref{ineq:gm-atmost-am} is because the geometric mean is at most the arithmetic mean, inequality \eqref{ineq:sqrt-e-bound} is because $\theta\ge 0$ and $\lambda^2+(1-\lambda)^2\ge 1/2$ for all $\lambda\in [0, 1]$, inequality \eqref{ineq:cheapest-to-lp} is because $a^u\in \conv(\calA_u)$ and $b^u\in \conv(\calB_u)$, and inequality \eqref{ineq:x-bound} is because $x_{\delta(u)}=\lambda_u a^u + (1-\lambda_u)b^u$. 
\end{proof}

%% file: bipartite.tex
\subsection{Bipartite Graphs}\label{sec:alg-bipartite}
Suppose the input graph is $G=(V, E)$ is bipartite. We pick $L$ to be one part of a valid bipartition of $G$ and use Algorithm \ref{algo:round}. Lemma \ref{lem:algo-correctness} implies that the solution returned by Algorithm \ref{algo:round} is indeed a feasible solution. 
We now analyze the approximation factor. 

\begin{lemma}\label{lem:bipartite-approximation}
    Let $(G, p,q, c)$ be a bipartite input instance of \FEC. 
    Let $F$ be the solution returned by Algorithm \ref{algo:round} implemented with $L$ being one part of a valid bipartition of $G$. Then, 
    \[
    E[c(F)]\le \left(1+\frac{1}{\sqrt{e}}\right)\sum_{e\in E}c_e x_e.
    \]
\end{lemma}
\begin{proof}
We observe that $F=F_1\uplus F_2$ and hence, $c(F)=c(F_1)+c(F_2)$. We bound the expected cost of $F_1$ and $F_2$ separately. 
First, we bound $E[c(F_1)]$. 
We have that 
\[
E[c(F_1)]=\sum_{v\in L}E[c(F_v)]=\sum_{v\in L}\sum_{e\in \delta(v)}c_e x_e = \sum_{e\in E}c_e x_e.
\]
The first and third equalities are owing to the bipartiteness of $G$.
The second equality is because for each $v\in L$,  for each edge $e\in \delta(v)$, the probability of sampling $e$ into $F_v$ is exactly $x_e$. 

Next, we bound $E[c(F_2)]$. Let $u\in V\setminus L$. For each $e\in \delta(u)$, the edge $e$ is sampled into $F_1$ \emph{independently} with probability $x_e$; independence holds because $G$ is a simple graph and the probability of sampling $e$ into $F_1$ depends only on the distribution at the other end-vertex $v\in L$ of $e$. Independence allows us to apply Lemma \ref{lem:augmentation-cost}. Hence, applying Lemma \ref{lem:augmentation-cost} with $\theta=1$, we have that 
\[
E[c(A_u)]\le \frac{1}{\sqrt{e}}\sum_{e\in \delta(u)}c_e x_e.
\]
Hence, 
\[
E[c(F_2)]=\sum_{u\in R}E[c(A_u)]\le\frac{1}{\sqrt{e}}\sum_{u\in R}\sum_{e\in \delta(u)}c_e x_e = \frac{1}{\sqrt{e}}\sum_{e\in E}c_e x_e.
\]
Again, the first and third equalities are owing to the bipartiteness of $G$. Consequently, $E[c(F)] = E[c(F_1) + c(F_2)]\le (1+1/\sqrt{e})\sum_{e\in E}c_e x_e$. 
\end{proof}

\begin{remark}
    We note that this algorithm cannot be de-randomized in the natural way; the natural procedure is for each vertex $v \in L$, compute a subset $F_{v}$ that has the cheapest $p+q$ edges incident to $v$ and for each vertex $u \in R$, compute the subset $A_{u}$. In the worst case, this procedure has an approximation factor asymptotically approaching two (see section \ref{sec:greedy-tight-example}).
\end{remark}

%% file: non-bipartite.tex
\subsection{Non-Bipartite Graphs}\label{sec:alg-arbitrary}
The choice of $L$ for bipartite graphs informs our rounding strategy for non-bipartite graphs. For bipartite graphs, we had the bipartition $L$ and $V\setminus L$ of vertices and our algorithm proceeded in two phases: we used the LP solution to sample and satisfy the covering requirements for vertices in $L$ in the first phase and in the second phase, we augmented the solution from the first phase with additional edges to meet the covering requirements for vertices in $V\setminus L$. For non-bipartite graphs, it is unclear how to pick the subset $L$ so that we can employ the same two-phase strategy. Two natural boundary cases are worth mentioning here: (i) if we choose to use the LP-solution to meet the covering requirements of vertices in $L=\emptyset$ in the first phase and augment to meet the covering requirements of vertices in $R=V$ in the second phase, then the resulting algorithm is simply the naive greedy algorithm described in the introduction which results in a $2$-factor which is also the best possible. 
(ii) if we choose to use the LP-solution to meet the covering requirements of vertices in $L=V$ in the first phase and augment to meet the covering requirements of vertices in $R=\emptyset$ in the second phase, then the resulting algorithm again achieves a $2$-factor (we were able to show this by adapting the analysis for bipartite graphs). 

For non-bipartite graphs, instead of deterministically deciding on $L$, we randomly sample each vertex $u\in V$ into the first phase set $L$ with probability $\theta$ (where the value of $\theta$ will be chosen later). 
For $\theta=0.47008$, we run Algorithm \ref{algo:round} for such a random choice of $L$. We now show that this rounding strategy achieves an approximation ratio of $1.78$.
\begin{lemma}\label{lem:General-approximation}
    Let $F$ be the solution returned by Algorithm \ref{algo:round} implemented for a random set $L$, where $L$ is obtained by picking each vertex $v\in V$ into $L$ with probability $\theta = 0.47008$. Then, 
    \[
    E[c(F)]\le 1.78\sum_{e\in E}c_e x_e.
    \]
\end{lemma}
\begin{proof}
    We observe that $F=F_1\uplus F_2$ and hence, $c(F)=c(F_1)+c(F_2)$. We bound the expected cost of $F_1$ and $F_2$ separately. 
First, we bound $E[c(F_1)]$. 
We have that 
\begin{align*}
E_{Z, F_1}[c(F_1)]
&=\sum_{v\in V}\pr[v\in L]E_{F_v}[c(F_v)|v\in L]\\
&=\sum_{v\in V}\theta E_{F_v}[c(F_v)|v\in L]\\
&= \sum_{v\in V}\theta \left(\sum_{e\in \delta(v)}c_e x_e\right) \\
&= 2\theta\left(\sum_{e\in E}c_e x_e\right).
\end{align*}
The second equality is because each $v\in V$ is sampled into $L$ with probability $\theta$. 
The third equality is because for each $v\in V$,  for each edge $e\in \delta(v)$, the probability of sampling $e$ into $F_v$ is exactly $x_e$. 
The final equality is because the $c_ex_e$ term for each $e\in E$ is counted from the perspective of both end-vertices of $e$. 

Next, we bound $E[c(F_2)]$. We note that 
\[
E_{Z, F_1}[c(F_2)]
=\sum_{u\in V}\pr[u\in V\setminus L]E_{F_1}[c(A_u)|u\in V\setminus L]. 
\]

Let $u\in V$. We have that $\pr[u\in V\setminus L]=1-\theta$. 
We now focus on $E_{F_1}[c(A_u)|u\in V\setminus L]$. Condition on the event that $u\in V\setminus L$. 
For each $e\in \delta(u)$, the edge $e$ is sampled into $F_1$ \emph{independently} with probability $\theta x_e$: 
This is because, the edge $e$ is present in $F_1$ only if the other end-vertex $v$ of $e$ is chosen into $L$ and conditioned on $v\in L$, we have that $e$ is sampled into $F_v$ with probability $x_e$. The probability of $v$ being sampled into $L$ is $\theta$ and conditioned on $v\in L$, the probability of $e\in F_v$ is $x_e$ (since $G$ is simple); thus, the probability of $e\in F_1$ is $\theta x_e$. Moreover, independence holds because $G$ is a simple graph. 
Independence allows us to apply Lemma \ref{lem:augmentation-cost}. Hence, applying Lemma \ref{lem:augmentation-cost} with $\theta$, we have that 
    \[E[c(A_u)|u\in V\setminus L]\le e^{-\frac{\theta}{2}}\sum_{e\in \delta(u)}c_e x_e.\]
Hence, 
\begin{align*}
E_{Z, F_1}[c(F_2)]
&=\sum_{u\in V}\pr[u\in V\setminus L]E_{F_1}[c(A_u)|u\in V\setminus L]\\
&=\sum_{u\in V}(1-\theta)E_{F_1}[c(A_u)|u\in V\setminus L]\\
&\le \sum_{u\in V}(1-\theta)e^{-\frac{\theta}{2}}\sum_{e\in \delta(u)}c_ex_e\\
&=2(1-\theta)e^{-\frac{\theta}{2}}\sum_{e\in E}c_ex_e. 
\end{align*}
The second inequality is because each $u\in V$ is in $V\setminus L$ with probability $1-\theta$. The final equality is because the $c_ex_e$ term for each $e\in E$ is counted from the perspective of both end-vertices of $e$. 

Consequently, we have that 
\begin{align*}
    E_{Z, F_1}[c(F_1)+c(F_2)]
    &\le 2\left(\theta + (1-\theta)e^{-\frac{\theta}{2}}\right) \sum_{e\in E}c_ex_e. 
\end{align*}
For $\theta=0.47001$, we have that $2 \theta + 2(1-\theta)e^{-\theta/2}=1.78$, thus bounding the approximation factor. 
Our choice of $\theta=0.47001$ is because the minimum value of the expression $f(\theta):=2 \theta + 2(1-\theta)e^{-\theta/2}$ over $\theta \in [0,1]$ occurs at $\theta^* = 0.47001$ with $f(\theta^*)=1.78$ (since $f$ is convex). 

\end{proof}

%% file: FGC_Disjunctive-2.tex
\section{Flexible Graph Connectivity via Disjunctive Programming}\label{sec:FGC}
In this section, we formulate a new LP-relaxation based on disjunctive programming for the flexible graph connectivity problem. This LP-relaxation is stronger than known relaxations given by Chekuri-Jain and Ibrahimpur-Vegh \cite{CJ2025, IV25} and is also polynomial-time solvable.
We recall the definition of \FGC. 
\begin{definition}
    In \FGC, the input is a graph $G = (V,E=S \uplus U)$, positive integers $p$ and $q$, and a cost function $c: E \rightarrow \mathbb{R}_{\geq 0}$. Here, the sets $S$ and $U$ are denoted as safe and unsafe edges, respectively. The goal is to find a subset $F \subseteq E$ with the least cost such that for every subset $\emptyset \neq X \subsetneq V$, $F$ has at least $p$ safe edges crossing the cut $X$ or $p+q$ edges crossing the cut $X$.
\end{definition}

This project started by examining Ibrahimpur-V\'{e}gh's LP relaxation for \FGC. We noticed that their LP is not integral for the special case where $G$ has only two vertices. This special case is a disjunctive variant of the Minimum-Cost-Unit-Weight-Knapsack problem, denoted \MCUW: 
    the input is a ground set $N = S \uplus U$, cost function $c: N \rightarrow \mathbb{R}_{\geq 0}$, and positive integers $p$ and $q$. The goal is to select a subset $F \subseteq N$ of minimum cost s.t. $|F \cap S| \geq p$ or $|F| \geq p+q$.
\MCUW can be solved in polynomial time as follows: Let $F_{S}$ be the $p$ cheapest elements of $S$ and $F$ be the $p+q$ cheapest elements of $N$. Output $\text{argmin}\left\{c(F_S),c(F)\right\}$. Our initial goal for a stronger LP for \FGC was to obtain an integral LP for this special case. We used the disjunctive programming approach to obtain such an LP. Extending this approach to arbitrary graphs, leads to the following LP-relaxation for \FGC.
\begin{align*}\tag{\textit{LP}$_3$}\label{eq:LP3}
        \min \quad&\sum_{e\in E} c_e x_e\\
        \quad& x_{e} = \alpha^{R}_{e} + \beta^{R}_{e} &\forall e \in \delta(v), \emptyset \neq R \subsetneq V \\
        & \alpha^{R}(\delta_{S}(R)) \geq p\lambda^{R} &\forall \emptyset \neq R \subsetneq V\\
        & \beta^{R}(\delta(R)) \geq (p+q)(1-\lambda^{R}) &\forall \emptyset \neq R \subsetneq V \\
        & 0 \leq \alpha^{R}_{e} \leq \lambda^{R} &\forall e \in \delta(R), \emptyset \neq R \subsetneq V\\
        & 0 \leq \beta^{R}_{e} \leq 1 - \lambda^{R} &\forall e \in  \delta(R), \emptyset \neq R \subsetneq V\\
        & 0 \leq \lambda^{R} \leq 1 &\forall \emptyset \neq R \subsetneq V
\end{align*}
By eliminating $\alpha$ and $\beta$ variables via Fourier-Motzkin elimination in a suitable order, we obtain the following LP that is equivalent to \eqref{eq:LP3}---we do not show the Fourier-Motzkin derivation in the interests of brevity. 
\begin{align*}\tag{\textit{LP}$_4$}\label{eq:LP_4}
    \min \quad&\sum_{e\in E}c_e x_e\\
    \quad& \sum_{e \in \delta_{S}(R)} \min \left\{x_e, \lambda^{R}\right\} \geq p \lambda^{R} &\forall \emptyset \neq R \subsetneq V \\
    & \sum_{e \in \delta(R)} \min \left\{x_e, 1- \lambda^{R}\right\} \geq (p+q)(1- \lambda^{R}) &\forall \emptyset \neq R \subsetneq V\\ 
    & \sum_{e \in \delta_{S}(R)} x_e + \sum_{e \in \delta_{U}(R)} \min \left\{x_e,1-\lambda^{R}\right\} \geq p \lambda^{R} + (p+q)(1-\lambda^{R}) &\forall \emptyset \neq R \subsetneq V\\
    & 0 \leq x_e \leq 1 &\forall e \in E\\
    & 0 \leq \lambda^{R} \leq 1  &\forall \emptyset \neq R \subsetneq V
\end{align*}

Next, we eliminate the $\lambda$ variables from \eqref{eq:LP_4} to arrive at the following LP: 
\begin{align*}\tag{\textit{LP}$_5$}\label{eq:LP_5}
    \min \quad\sum_{e\in E}c_e x_e\\
    x(\delta_{S}(R)) + x(\delta_{U}(R) - J) &\geq p \quad \quad \quad \quad \quad \quad \quad \quad \quad \quad \forall J \subseteq \delta_{U}(R): |J| \le q \\ 
    \frac{x(\delta_S(R)-T)}{p-|T|} + \frac{x(\delta_S(R)) + x(\delta_U(R) - J) - p}{q-|J|} &\geq 1 \quad 
    \forall T \subseteq \delta_S(R): |T| < p, \,\ \forall J \subseteq \delta_U(R): |J| < q
    \\
    0 \leq x_e &\leq 1 \quad \quad \quad \quad \quad \quad \quad \quad \quad \quad \quad \quad \quad \quad  \forall e \in E
\end{align*}
We briefly outline why \eqref{eq:LP_4} and \eqref{eq:LP_5} are equivalent. If we have a feasible solution $x$ for \eqref{eq:LP_5}, then the solution $(x, (\lambda^R)_{\emptyset\neq R\subsetneq V})$ obtained by setting $\lambda^R:=\min\{a^R, 1-b^R\}$ for each $\emptyset\neq R\subsetneq V$, where 
\begin{align*}
    a^R&:=\min_{T\subseteq \delta_S(R): |T|<p}\frac{x(\delta_S(R)\setminus T)}{p-|T|}, \text{ and}\\
    b^R&:=
    \begin{cases}
        q\text{'th largest }x_e \text{ value among edges }e\in \delta_U(R) & \text{ if }|\delta_U(R)|\ge q\\
        0 &\text{ otherwise}.
    \end{cases}
\end{align*}
is feasible for \eqref{eq:LP_4}. Moreover, if $(x, (\lambda^R)_{\emptyset\neq R\subsetneq V})$ is a feasible solution for \eqref{eq:LP_4}, then $x$ can be shown to satisfy the constraints of \eqref{eq:LP_5} by a straightforward derivation. We do not elaborate on these details since these LPs are not the main focus of this work. 

\paragraph{Solvability of the LP.} We note that \eqref{eq:LP_5} can be solved in polynomial time via the Ellipsoid method since it admits a polynomial time separation oracle (even if $p$ and $q$ are part of input and not fixed constants). 
For each family of constraints, the separation problem can be reduced to cardinality-constrained global min-cut interdiction. Zenklusen \cite{Z2014} showed that cardinality-constrained global min-cut interdiction can be solved in polynomial time via reduction to $2$-criteria global minimum cut which, in turn, is known to be polynomial-time solvable \cite{AZ06}. 

\paragraph{Strength of the LP compared to known LPs.} 
\eqref{eq:LP_5} dominates the relaxation of Ibrahimpur-V\'{e}gh \cite{IV25}, which in turn dominates that of Chekuri-Jain \cite{CJ2025}. The first dominance can be strict, even on simple graphs. Indeed, the first family of constraints in \eqref{eq:LP_5}, together with the second family restricted to $T=J=\emptyset$, recovers the Chekuri--Jain relaxation. The full second family strengthens the corresponding Ibrahimpur-V\'{e}gh inequalities, while the remaining inequalities of their relaxation follow from the first family and the box constraints. Beyond this strengthening, the disjunctive formulation exposes local convex decompositions that underpin our rounding algorithms for \FEC. These results suggest that exploiting the same structure across general cuts is a promising route toward a constant-factor algorithm for \FGC.

%% file: appendix.tex
\section{Appendix}\label{sec:appendix}
\subsection{Integrality Gap of \eqref{eq:LP1}}\label{sec:lp1-integ-gap}
We show that \eqref{eq:LP1} has unbounded integrality gap. 
\begin{proposition}
        \eqref{eq:LP1} has unbounded integrality gap on simple bipartite graphs.
    \end{proposition}
    \begin{proof}
    We construct a simple bipartite instance to exhibit an unbounded integrality gap. 
        Let $p \geq 2$ and $G = K_{p+q,p+q}$ be the complete bipartite graph on $p+q$ vertices on both sides. Let the bipartition be $V = L \sqcup R$. We assign the edges in $G$ to $S$ and $U$ as follows. Decompose $G$ into $p+q$ edge disjoint perfect matchings $\left\{M_1,M_2,\ldots,M_{p+q}\right\}$ (this is possible because $G$ is a complete bipartite graph with equal number of vertices on both sides). Let $S = M_{p+q} \cup \bigcup_{i=1}^{p-1} M_i$ and $U = \bigcup_{i \in \left\{p,p+1,\cdots,p+q-1\right\}} M_i$. Define the cost function $c: E \rightarrow \mathbb{R}_{\geq 0}$ as
        \[
        c_{e} = 
        \begin{cases}
            \epsilon \quad \text{if } e \in \bigcup_{i \in [p+q-1]} M_i\\
            1 \quad \text{otherwise}
        \end{cases}
        \]
        Let $F$ be an optimal integral solution to this instance of $\FEC$. Since $|\delta(v)| = p+q$ and $|\delta_{S}(v)| = p$, $F \cap \delta(v)$ must include the edge $e_{v} \in M_{p+q} \cap \delta(v)$. Hence, $c(F) \geq \sum_{v \in L} c_{e_{v}} = p + q$. We now construct a feasible solution $(x,\lambda)$ to \ref{eq:LP1}. Let $\lambda_{v} = \frac{1}{p+q}$ for all vertices $v \in V$ and let
        \[
        x_{e}=
        \begin{cases}
            1 \quad \text{if } e \in \bigcup_{i \in [p+q-1]} M_i\\
            0 \quad \text{otherwise}.
        \end{cases}
        \]
        Then, $0 \leq x \leq 1$ and $0 \leq \lambda \leq 1$. For every vertex $v \in V$ 
        \[x(\delta(v)) = p+q-1 \geq 1 = \lambda_{v} (p+q)\]
        \[x(\delta_{S}(v)) = p \geq \frac{p}{p+q} = \lambda_{v} p.\]
        Thus $(x,\lambda)$ is feasible for \ref{eq:LP1} and the integrality gap is at least 
        \[\frac{c(F)}{\sum_{e \in E} c_e x_{e}} \geq \frac{p+q}{\epsilon(p+q)(p+q-1)} = \frac{1}{\epsilon (p+q-1)}\]
        which tends to $\infty$ as $\epsilon \rightarrow 0$. 
    \end{proof}

\subsection{Tight example for the naive greedy heuristic}\label{sec:greedy-tight-example}
In this section, we investigate the naive greedy heuristic stated in the introduction: namely, for each vertex $v \in V$,  let $F_{v}^1$ be the cheapest $p+q$ edges incident to $v$, $F_v^2$ be the cheapest $p$ safe edges incident to $v$, and $F_v=\arg\min\{c(F_v^1), c(F_v^2)\}$; return $F = \bigcup_{v \in V} F_{v}$. We construct an instance to illustrate that $2$-factor is the best possible for this heuristic. 

Let $p = 1$ and $q=0$. We construct a graph $G = (V = L \sqcup R, E=S)$ where $L = \left\{\ell_1,\ell_2,\cdots, \ell_n\right\}$, $R = \left\{r_1,r_2,\cdots, r_n\right\}$, $E = \left\{\ell_ir_i \mid i \in [n]\right\} \cup \left\{\ell_i r_1 \mid i \in [n] \setminus \left\{1\right\}\right\}$. Define a cost function $c: E \rightarrow \mathbb{R}_{\geq 0}$ where $c_{\ell_i r_i} = 1 + \epsilon$ for every $i \geq 1$ and  $c_{\ell_i r_1} = 1$ for every $i \geq 2$. For $\epsilon$ sufficiently small, the optimal solution $F^{*}$ is $F^{*} = \left\{\ell_ir_i \mid i \in [n]\right\}$ with $c(F^{*}) = n(1+\epsilon)$. We can adversarially set the processing order of the vertices to be $L$ first and then $R$. Then, the algorithm will construct the solution $F = E$. We have that $\frac{c(F)}{c(F^{*})} = \frac{n(1+\epsilon) + n-1}{n(1+\epsilon)} = 1 + \frac{n-1}{n(1+\epsilon)}$. As $n \rightarrow \infty$ and $\epsilon \rightarrow 0$, this ratio approaches $2$. We note that this example is a simple graph.